\documentclass[showpacs,twocolumn,prl,aps]{revtex4}
\usepackage{graphicx}

\begin{document}

\title[Short title for running header]{Topological entanglement entropy in Gutzwiller projected spin liquids}
\author{Jiquan Pei, Steve Han, Haijun Liao and  Tao Li}
\affiliation{ Department of Physics, Renmin University of China,
Beijing 100872, P.R.China}
\date{\today}

\begin{abstract}
The topological entanglement entropy(TEE) of Gutzwiller projected
RVB state is studied with Monte Carlo simulation. New tricks are
proposed to improve the convergence of TEE, which enable us to show
that the spin liquid state studied in Ref.\cite{Vishwanath} actually
does not support $Z_{2}$ topological order, a conclusion that is
consistent with the information drawn from the inspection of the
topological degeneracy on the same state. We find both a long ranged
RVB amplitude and an approximate Marshall sign structure are at the
origin of the suppression of vison gap in this spin liquid state. On
the other hand, robust signature of $Z_{2}$ topological order, i.e.,
a TEE of $\ln2$, is clearly demonstrated for a Gutzwiiler projected
RVB state on triangular lattice which is evolved from the RVB state
proposed originally by Anderson\cite{Sorella}. We also find that it
is the sign, rather than the amplitude of the RVB wave function,
that dominates the TEE and is responsible for a positive value of
TEE, which implies that the nonlocal entanglement in the RVB state
is mainly encoded in the sign of the RVB wave function. Our results
indicate that some information that is important for the topological
property of a RVB state is missed in the effective field theory
description and that a $Z_{2}$ gauge structure in the saddle point
action is not enough for the RVB state to exhibit topological order,
even if its spin correlation is extremely short ranged.
\end{abstract}
 \maketitle

\section{I.Introduction}
The search for spin liquid phases in quantum antiferromaget is a hot
topic in the field of strongly correlated electron
systems\cite{Frustrated}. A spin liquid state is an exotic state of
matter in that it can exhibit novel structures and excitations
beyond the Landau-Ginzburg paradigm. The topological order and the
related fractionalized excitation are one of such possibilities. A
topological ordered system is unique in that its ground state
exhibits topological degeneracy on multiply connected manifolds. A
topological ordered state is also associated with a nonzero
topological contribution to the entanglement
entropy\cite{Kitaev,Levin}.

Following the original idea of resonating valence bond state, two
approaches are widely used for the description of spin liquid phase.
The first one is the effective field theory approach based on slave
particle representation of the spin degree of freedom\cite{Lee}. The
second one is the variational approach based on Gutzwiller projected
wave functions\cite{Sorella}. The Gutzwiller projection procedure,
which removes the doubly occupied configuration in the mean field
ground state, amounts to perform gauge averaging on the saddle point
of the effective field theory. Thus, if the gauge fluctuation is
irrelevant in the long wavelength limit, one should expect the
Gutzwiller projected RVB state to exhibit the same long range
physics as the effective field theory. In a gapped spin liquid, the
only nontrivial physics in the long wavelength limit is its
topological property. However, it is found earlier that the
topological property predicted by the effective field theory does
not always agree with that found from the corresponding Gutzwiller
projected wave function. In particular, the Marshall sign rule is
found to be closely related to the the absence of the topological
order in certain spin liquid states\cite{Li1,Li2,Li3,Sorella1}.

Both the topological degeneracy and the topological entanglement
entropy(TEE) are used to characterize the topological property of a
spin liquid state. To detect the topological degeneracy, wave
functions with different number of trapped visons in the holes of a
multiply connected manifold are constructed\cite{Ivanov,Li1,Arun}.
The overlap of these wave functions, which should decay
exponentially with the linear scale of the system, provides a
measure of the size of the vison gap\cite{Arun}. On the other hand,
only a single ground state is needed to extract the
TEE\cite{Furukawa,Hastings,Jiang}. The TEE of the Gutzwiller
projected spin liquid state is first studied in
Ref.\cite{Vishwanath,errutum} with a Monte Carlo sampling procedure.
More specifically, they evaluated the TEE of a Gutzwiller projected
spin liquid state derived from a mean field ansatz with a $Z_{2}$
gauge structure and a very large spin gap, hereafter denoted as
state RVB-I, and claimed that the obtained TEE($0.584\pm 0.089$) is
consistent with the expectation for a $Z_{2}$ spin liquid(for which
$\gamma=\ln2\approx 0.693$). However, as a result of the poor
convergence the true error bar is much larger than claimed. A latter
calculation with higher accuracy by the same group shows that the
TEE of state RVB-I is far away($0.288\pm0.107$) from the expected
value of $\ln2$\cite{errutum}.

In this paper, new tricks are proposed to improve the convergence of
the TEE of the Gutzwiller projected wave function in Monte Carlo
sampling. With these tricks significant reduction in the error bar
can be achieved and calculation on larger subsystem is possible. We
find the TEE of the state RVB-I is about $0.2\pm0.04$ for the
largest subsystem size we have tried, which is significantly smaller
than $\ln2$. The TEE is also found to have negligible dependence on
the size of spin gap. Both of these results indicate that the
topological property of the state RVB-I is different from the
prediction of the effective field theory.

As a crosscheck we have made a test of topological degeneracy on the
state RVB-I. We find the vison gap of this spin liquid state is
extremely small, if it does exist at all. We find a long ranged RVB
amplitude(even though the spin correlation is extremely short ranged
in the state RVB-I) and an approximate Marshall sign structure are
responsible for the suppression of the vison gap in the state RVB-I.
We also find that it is the sign of the RVB wave function, rather
than its amplitude, that dominates the TEE and is responsible for a
positive value of TEE in this state. The reduction of TEE in the
state RVB-I can thus be understood as a direct consequence of the
approximate Marshall sign structure in this state.

With these understandings, we have evaluated the TEE of a Gutzwiller
projected spin liquid state on triangular lattice with a short
ranged RVB amplitude and a frustrated sign. This RVB state can be
connected smoothly to the short range RVB state first proposed by
Anderson\cite{Anderson,Sorella} and will be denoted as RVB-II
hereafter. We find the TEE of this state converges steadily to the
expected value of $\ln2$. Again we find it is the sign of the wave
function that is responsible for a positive TEE. These results
indicate that some information that is important for the topological
property of a RVB state is missed in the effective field theory
description and that the nonlocal entanglement in the RVB state is
mainly encoded in the sign of the RVB wave function.

This paper is organized as follows. In the next section, we
introduce the Gutzwiller projected RVB state and the related
effective field theory description. We then outline the basic steps
to evaluate the Renyi entanglement entropy and the topological
entanglement entropy of such states. In section III, we introduce
new tricks to improve the convergence of the TEE. In section IV, we
apply these tricks to study the topological properties of the spin
liquid state RVB-I. In the same section we also make a test of
topological degeneracy on the state RVB-I and discuss the reason for
the suppression of vison gap in this state. In section V, we present
the result of TEE for RVB-II and discuss the relation between the
sign structure of the RVB wave function and the TEE. In the last
section, we present some discussions on the relation between the
effective field theory and the physics of RVB state.

\section{II. The Gutzwiller projected RVB state and its topological entanglement entropy}

The Gutzwiller projected BCS mean field ground state is widely used
to describe a spin liquid state. It has the following general form
\begin{equation}
|\Psi\rangle=P_{\mathrm{G}}|\mathrm{BCS}\rangle.
\end{equation}
Here $|\mathrm{BCS}\rangle$ is the ground state of the following BCS
mean field Hamiltonian
\begin{equation}
\mathrm{H}_{\mathrm{BCS}}=-\sum_{\langle
i,j\rangle}\psi_{i}^{\dagger}U_{i,j}\psi_{j},
\end{equation}
in which $\psi_{i}=\left(\begin{array}{c}
                           c_{i\uparrow} \\
                           c^{\dagger}_{i\downarrow}
                         \end{array}
\right)$ is the Nambu spinor. $U_{ij}=\left(\begin{array}{cc}
                   -\chi^{*}_{i,j} & \Delta_{i,j} \\
                   \Delta^{*}_{i,j} & \chi_{i,j}
                 \end{array}\right)
$ is a $2 \times 2$ matrix denoting the hopping and pairing order
parameters of $\psi_{i}$. On the other hand, this mean field
Hamiltonian can be viewed as a saddle point approximation of a spin
liquid state in the effective field theory description based on
slave Boson representation of the spin. The fluctuation of the order
parameter $U_{ij}$ around their saddle point value takes the form of
gauge fluctuation in the low energy regime. Depending on the
structure of $U_{i,j}$, the gauge fluctuation can be either gapped
or gapless in the long wavelength limit. In particular, when the
gauge symmetry is broken down to $Z_{2}$ in the saddle point, the
gauge fluctuation is believed to be gapped and the low energy
physics of the spin liquid state is believed to be faithfully
represented by the saddle point approximation\cite{Lee}. In such a
case, the only effect of the gauge degree of freedom in the long
wavelength limit is to induce a $Z_{2}$ topological degeneracy on
multiply connected manifold.

From the point of view of the effective field theory, the Gutzwiller
projection procedure amounts to perform gauge averaging on the
saddle point. Thus, when the gauge fluctuation is gapped we should
expect the Gutzwiller projected RVB state to exhibit the same long
wavelength behavior as the corresponding effective field theory. In
particular, the topological property predicted by both approaches
should be the same.

The topological property of the RVB state can be characterized by
the topological entanglement entropy. To define the entanglement
entropy on a RVB state, we partition the system into a subsystem and
the corresponding environment. The reduced density matrix of the
subsystem is then given by tracing out the degree of freedom in the
environment
\begin{equation}
\rho_{s}=\mathrm{Tr}_{e} | \Psi \rangle \langle \Psi |,
\end{equation}
and the entanglement entropy of the Von Neumann type is defined as
\begin{equation}
S=-\mathrm{Tr}\rho_{s}\ln\rho_{s}.
\end{equation}
The entanglement entropy in general satisfies the area law $S=\alpha
L-\gamma$, in which $L$ is the length of the boundary separating the
subsystem from the environment. $\gamma$ is the universal
topological contribution and is nonzero in a topological ordered
state.

It is in general quite hard to evaluate the Von Neumann entropy from
the definition. However, it is proved that the TEE can also be
extracted from the more general Renyi entropy\cite{Flammia}. The
Renyi entropy of order $n$ is defined as
\begin{equation}
S_{n}=-\frac{1}{n-1}\ln \mathrm{Tr}\rho_{s}^{n}.
\end{equation}
The Von Neumann entropy is related to the Renyi entropy by
$S=\lim_{n\rightarrow 1}S_{n}$. It is proved that the topological
contribution to the Renyi entanglement entropy is independent of the
value of n\cite{Flammia}.

In the following, we will use $S_{2}$ to extract the TEE. To
evaluate $S_{2}$, we use the replica method and introduce an
identical copy of the system\cite{Hastings,Vishwanath}. The replica
is partitioned in exact the same manner as the system into a
subsystem and an environment. Denoting the state vector of the
system and its replica as a whole as $|\Psi\otimes\Psi\rangle$, it
can shown that
\begin{equation}
\mathrm{Tr}\rho_{s}^{2}=\langle\Psi \otimes \Psi |\mathrm{SWAP}|\Psi
\otimes \Psi\rangle.
\end{equation}
Here $\mathrm{SWAP}$ denotes the operation of exchanging all the
degree of freedoms within the subsystem between the system and its
replica\cite{Hastings}. To evaluate $\mathrm{Tr}\rho_{s}^{2}$, we
introduce an orthonormal basis $\{s_{i}\}$ in both the system and
its replica and denote the wave function in this basis as
$\psi(s_{i})$. Then we have
\begin{equation}
\mathrm{Tr}\rho_{s}^{2}=\frac{\sum_{s_{i},s_{i'}}|\psi(s_{i})|^{2}|\psi(s_{i'})|^{2}\frac{\psi(\tilde{s}_{i})\psi(\tilde{s}_{i'})}{\psi(s_{i})\psi(s_{i'})}}{\sum_{s_{i},s_{i'}}|\psi(s_{i})|^{2}|\psi(s_{i'})|^{2}},
\end{equation}
here $s_{i}$ and $s_{i'}$ are the spin configurations of the system
and its replica before the swap operation. $\tilde{s}_{i}$ and
$\tilde{s}_{i'}$ are the swapped spin configurations in the system
and its replica. In Ref.\cite{Vishwanath}, a sign trick is proposed
to evaluate the above average. For this purpose we rewrite the above
average as follows
\begin{equation}
\langle\mathrm{SWAP}\rangle=\langle\mathrm{SWAP_{amp}}\rangle\times\langle\mathrm{SWAP_{sign}}\rangle,
\end{equation}
in which
\begin{eqnarray}
\langle\mathrm{SWAP_{amp}}\rangle&=&\frac{\sum_{s_{i},s_{i'}}\rho(s_{i},s_{i'})|\frac{\psi(\tilde{s}_{i})\psi(\tilde{s}_{i'})}{\psi(s_{i})\psi(s_{i'})}|}{\sum_{s_{i},s_{i'}}\rho(s_{i},s_{i'})}
\nonumber\\
\langle\mathrm{SWAP_{sign}}\rangle&=&\frac{\sum_{s_{i},s_{i'}}\tilde{\rho}(s_{i},s_{i'})e^{i\phi(s_{i},s_{i'})}}{\sum_{s_{i},s_{i'}}\tilde{\rho}(s_{i},s_{i'})}.
\end{eqnarray}
Here $\rho(s_{i},s_{i'})=|\psi(s_{i})|^{2}|\psi(s_{i'})|^{2}$,
$\tilde{\rho}(s_{i},s_{i'})=|\psi(\tilde{s}_{i})\psi(\tilde{s}_{i'})\psi(s_{i})\psi(s_{i'})|$.
$\phi(s_{i},s_{i'})$ is the phase of
$\frac{\psi(\tilde{s}_{i})\psi(\tilde{s}_{i'})}{\psi(s_{i})\psi(s_{i'})}$.
For both the spin liquid state RVB-I and RVB-II studied in this
work, it can be shown that the wave function $\psi(s_{i})$ is real
up to a global phase as a result of the time reversal symmetry. Thus
$e^{i\phi(s_{i},s_{i'})}=\pm 1$.

In general, we expect both $\langle\mathrm{SWAP_{sign}}\rangle$ and
$\langle\mathrm{SWAP_{amp}}\rangle$ to decrease exponentially with
the subsystem size, as is required by the area law for the
entanglement entropy. Following Ref.\cite{Vishwanath1}, we define
$S^{\mathrm{sign}}_{2}$ and $S^{\mathrm{amp}}_{2}$ as the
contribution of the phase and the amplitude of the wave function to
the entanglement entropy. They are given by
\begin{eqnarray}
S^{\mathrm{sign}}_{2}&=&-\ln\langle \mathrm{SWAP_{sign}} \rangle\nonumber\\
S^{\mathrm{amp}}_{2}&=&-\ln\langle \mathrm{SWAP_{amp}} \rangle
\end{eqnarray}

\begin{figure}[h!]
\includegraphics[width=8cm,angle=0]{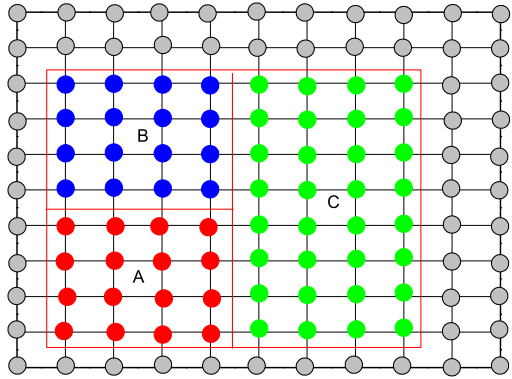}
\caption{An illustration of the partition of the degree of freedoms
on the square lattice used to extract the TEE from Eq.(11).}
\label{fig1}
\end{figure}

In principle, the TEE can be extracted from the defining equation
$S_{2}=\alpha L-\gamma$ by an appropriate extrapolation procedure.
However, the result thus obtained suffers from the ambiguity in the
definition of $L$ on a lattice. To remove such ambiguity, a bulk and
boundary cancelation procedure is widely used. As in
Ref.\cite{Vishwanath}, we divide the subsystem into three parts, A,
B and C. An illustration of the partition on the square lattice is
shown in Fig.1. Then the topological contribution to the
entanglement entropy can be extracted by subtracting out both the
the bulk and boundary contributions from the following combination
\begin{equation}
-\gamma=S_{2}^{\mathrm{A}}+S_{2}^{\mathrm{B}}+S_{2}^{\mathrm{C}}-S_{2}^{\mathrm{AB}}-S_{2}^{\mathrm{AC}}-S_{2}^{\mathrm{BC}}+S_{2}^{\mathrm{ABC}}.
\end{equation}

\section{III. The trick of re-weighting and the trick of ratio estimator}

The evaluation of TEE involves the computation of the sign
contribution $S^{\mathrm{sign}}_{2}$ and amplitude contribution
$S^{\mathrm{amp}}_{2}$ for each of the seven subsystems in Eq.(11).
For a general RVB state, both $S^{\mathrm{sign}}_{2}$ and
$S^{\mathrm{amp}}_{2}$ follows the area law. However, the
logarithmic correction to the area law in a gapless spin liquid
state is found to be mainly contributed by
$S^{\mathrm{sign}}_{2}$\cite{Vishwanath1}.

As we will see below, for the state RVB-I, $S_{2}$ is mainly
contributed by $S^{\mathrm{amp}}_{2}$ as a result of an approximate
Marshall sign rule. For large subsystem size, the convergence of
$S^{\mathrm{amp}}_{2}$ becomes very slow. This is expected since the
spin configurations before and after the swap operation become
increasingly remote from each other in the Hilbert space with the
increase of the subsystem size. As a result, the fluctuation in the
quantity to be averaged,
$R=\frac{\psi(\tilde{s}_{i})\psi(\tilde{s}_{i'})}{\psi(s_{i})\psi(s_{i'})}$,
grows rapidly with the subsystem size. However, the averaged value
of such a ratio should decrease exponentially with the length of the
boundary, as is required by the area law. A Monte Carlo sampling of
such a wildly fluctuating quantity is thus very hard and this
explains why Ref.\cite{Vishwanath} fails to reach the true
convergence in the value of TEE.

In the following, we propose two tricks to improve the convergence
of the $S^{\mathrm{amp}}_{2}$. The first trick is a simple
re-weighting procedure. Rather than sampling the weight
$W(s_{i},s_{i'})=\rho(s_{i},s_{i'})$, we sample the combined weight
$W'(s_{i},s_{i'})=\rho(s_{i},s_{i'})+\rho(\tilde{s}_{i},\tilde{s}_{i'})$.
It is shown in Ref.\cite{Li4} that such a combined sampling can
reduce the statistical fluctuation. With such a combined weight, the
quantity to be averaged becomes
\begin{equation}
\langle\mathrm{SWAP_{amp}}\rangle=2\frac{\sum_{s_{i},s_{i'}}W'(s_{i},s_{i'})\frac{R}{1+R^{2}}}{\sum_{s_{i},s_{i'}}W'(s_{i},s_{i'})}.
\end{equation}
Now the quantity to be averaged becomes $R/(1+R^{2})$, whose
fluctuation is the square root of $R$.

For even larger subsystem size, the above re-weighting trick will
also fail as the fluctuation in $R$ grows exponentially with the
subsystem size. A trick to cope with such an exponentially growing
fluctuation is the ratio estimator\cite{Hastings,Pei}. In such a
trick, we express the quantity to be evaluated as the product of a
series of ratios so that the evaluation of each ratio only suffers
from a much smaller fluctuation. There are different ways to
implement this trick. One convenient way is to rewrite
$\langle\mathrm{SWAP_{amp}}\rangle$ in the following form
\begin{equation}
\langle\mathrm{SWAP_{amp}}\rangle=\frac{\sum_{s_{i},s_{i'}}A}{\sum_{s_{i},s_{i'}}B}=\prod_{i=1}^{m}\frac{\sum_{s_{i},s_{i'}}A^{1-r_{i}}B^{r_{i}}}{\sum_{s_{i},s_{i'}}A^{1-r_{i+1}}B^{r_{i+1}}},
\end{equation}
in which
$A=|\psi(\tilde{s}_{i})\psi(\tilde{s}_{i'})\psi(s_{i})\psi(s_{i'})|$
, $B=|\psi(s_{i})|^{2}|\psi(s_{i'})|^{2}$. $r_{i}\in[ 0,1 ]$ is a
series of powers satisfying $r_{i}<r_{i+1}$, $r_{1}=0$ and
$r_{m+1}=1$. If we define
$\tilde{W}_{i}(s_{i},s_{i'})=A^{1-r_{i+1}}B^{r_{i+1}}$, then it can
be shown that
\begin{equation}
\langle\mathrm{SWAP_{amp}}\rangle=\prod_{i=1}^{m}\frac{\sum_{s_{i},s_{i'}}\tilde{W}_{i}(s_{i},s_{i'})(R)^{r_{i+1}-r_{i}}}{\sum_{s_{i},s_{i'}}\tilde{W}_{i}(s_{i},s_{i'})}.
\end{equation}
When $r_{i+1}-r_{i}$ are chosen to be sufficiently small, each term
in the above product can be evaluated easily. We note a similar
trick is recently proposed to compute the TEE of quantum dimer model
at the R-K point by the present authors\cite{Pei}.

\section{IV. The topological properties of the state RVB-I}

In Ref.\cite{Vishwanath}, the TEE of the spin liquid state with the
following $Z_{2}$ mean field ansatz is studied,
\begin{eqnarray}
U_{i,i+x}&=&U_{i,i+y}=-\tau^{3}\nonumber\\
U_{i,i+x+y}&=&\eta(\tau^{1}+\tau^{2})\nonumber\\
U_{i,i-x+y}&=&\eta(\tau^{1}-\tau^{2})\nonumber\\
U_{i,i}&=&\lambda\tau^{1},
\end{eqnarray}
in which $\tau^{1,2,3}$ are the three Pauli matrices, $\eta$ and
$\lambda$ are two real parameters. $\lambda$ is determined self
consistently from the mean field condition of $\langle
c_{i\downarrow}c_{i\uparrow}\rangle=0$. The Gutzwiller projected RVB
state derived from this ansatz describes a symmetric spin liquid.
The RVB amplitude is given by
\begin{equation}
a(\mathrm{R}_{i}-\mathrm{R}_{j})=\sum_{\mathrm{k}}\frac{\Delta_{\mathrm{k}}}{\xi_{\mathrm{k}}+\sqrt{\xi_{\mathrm{k}}^{2}+|\Delta_{\mathrm{k}}|^{2}}}e^{i\mathrm{k}\cdot(\mathrm{R}_{i}-\mathrm{R}_{j})},
\end{equation}
in which
$\xi_{\mathrm{k}}=-2(\cos(\mathrm{k}_{x})+\cos(\mathrm{k}_{x}))$,
$\Delta_{\mathrm{k}}=\lambda+4\eta\cos(\mathrm{k}_{x})\cos(\mathrm{k}_{y})+4i\eta\sin(\mathrm{k}_{x})\sin(\mathrm{k}_{y})$.
At the effective field theory level, this spin liquid phase is
predicted to possess $Z_{2}$ topological order since the gauge
symmetry is broken down to $Z_{2}$ at the saddle point and that spin
excitation spectrum is fully gapped. We thus expect to see a TEE of
$\ln2$ in the state RVB-I.

\begin{figure}[h!]
\includegraphics[width=8cm,angle=0]{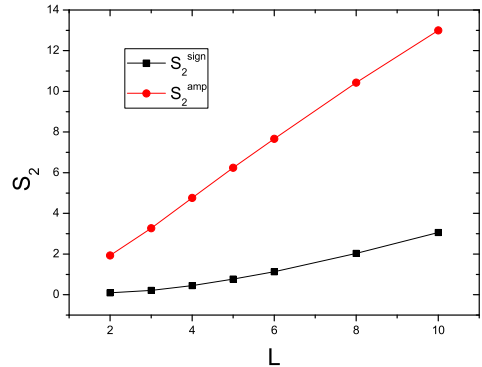}
\caption{The Renyi entanglement entropy from the amplitude and the
sign contribution in the state RVB-I. The subsystem is a $L\times L$
square region and the total system size is $16\times16$. }
\label{fig2}
\end{figure}
In our calculation of the entanglement entropy, we find the sign
contribution $S^{\mathrm{sign}}_{2}$ for this state is much smaller
than the amplitude contribution $S^{\mathrm{amp}}_{2}$, as is
illustrated in Fig.2. The origin for such an unusual behavior is an
approximate Marshall sign structure in this RVB state. Although the
ansatz Eq.(15) is frustrated and the Marshall sign rule is in a
strict sense violated, we find the extent to which the Marshall sign
rule is violated is always limited. To illustrate this point, we
plot in Fig.3 the average of the Marshall sign in the state RVB-I.
The averaged Marshall sign is defined as
$\langle(-1)^{N^{\downarrow}_{A}}\rangle$\cite{Sorella2}, in which
$N^{\downarrow}_{A}$ is the number of down spins in sublattice $A$.
\begin{figure}[h!]
\includegraphics[width=8cm,angle=0]{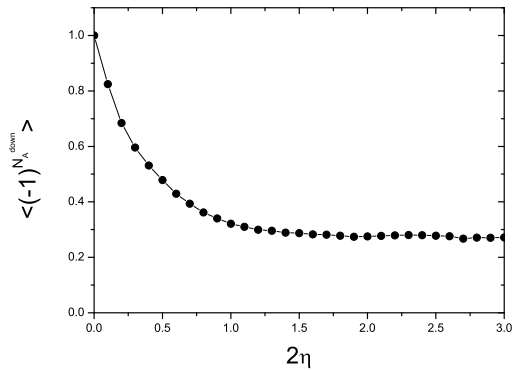}
\caption{The averaged Marshall sign in the spin liquid state RVB-I
as a function of the gap parameter $\eta$. } \label{fig3}
\end{figure}

As a result of such a sign rule, $S^{\mathrm{sign}}_{2}$ converges
rapidly. However, the evaluation of $S^{\mathrm{amp}}_{2}$ is very
challenging. In the evaluation of $S^{\mathrm{amp}}_{2}$, we have
used both the re-weighting and the ratio estimator trick introduced
in the last section. In Fig.4a, we show the result of TEE for
various subsystem sizes. The TEE is seen to be much smaller than the
value of $\ln2$ predicted by the $Z_{2}$ effective field theory. To
test if the deviation from the effective field theory prediction is
caused by the finite size effect in spin correlation, we have
studied the dependence of TEE on the size of the spin gap, the
result of which is shown in Fig.4b. The TEE is found to be
insensitive to the size of spin gap. Thus it is unlikely that the
deviation from effective field theory prediction is caused by a
finite size effect in the spin correlation.
\begin{figure}[h!]
\includegraphics[width=8cm,angle=0]{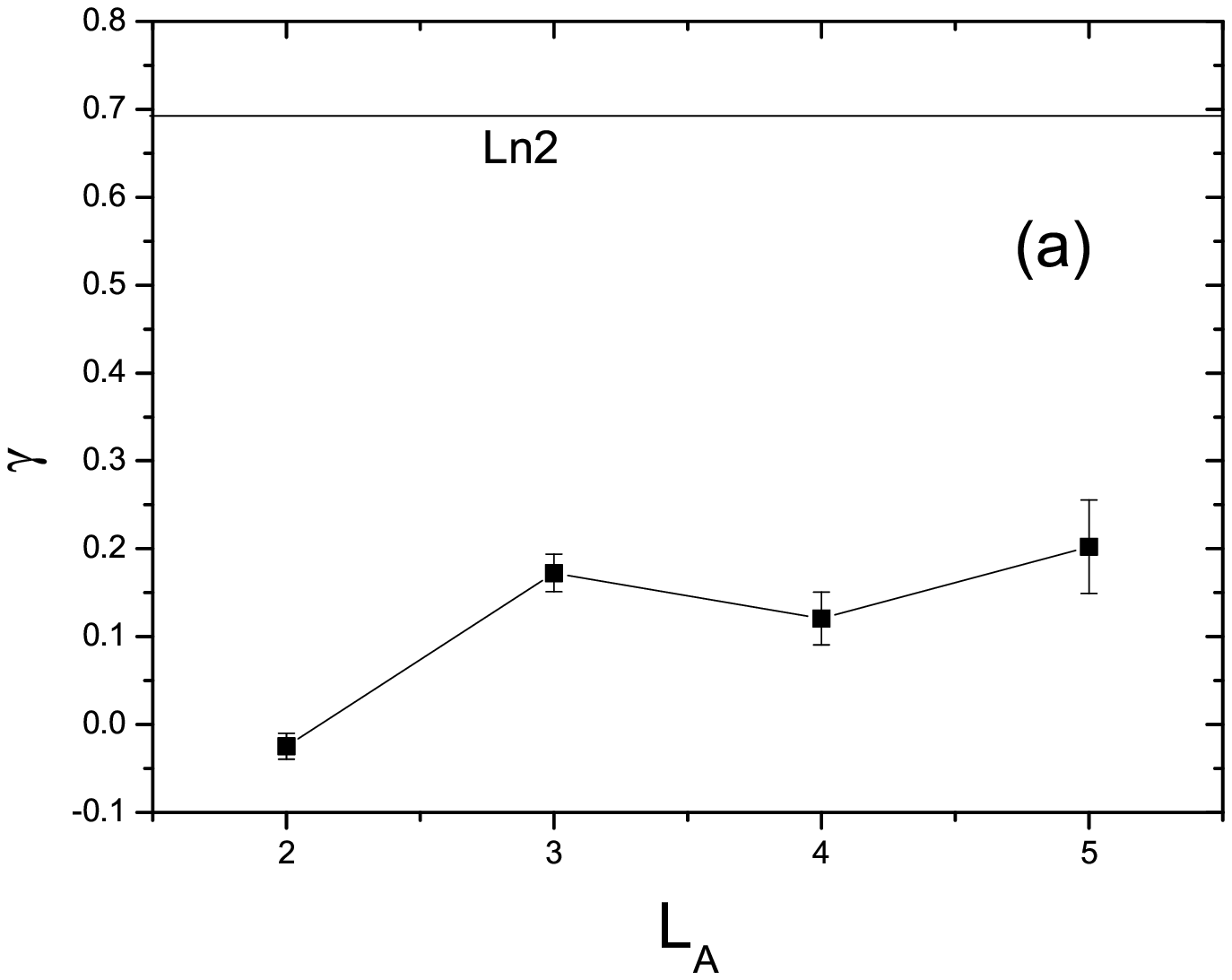}
\includegraphics[width=8cm,angle=0]{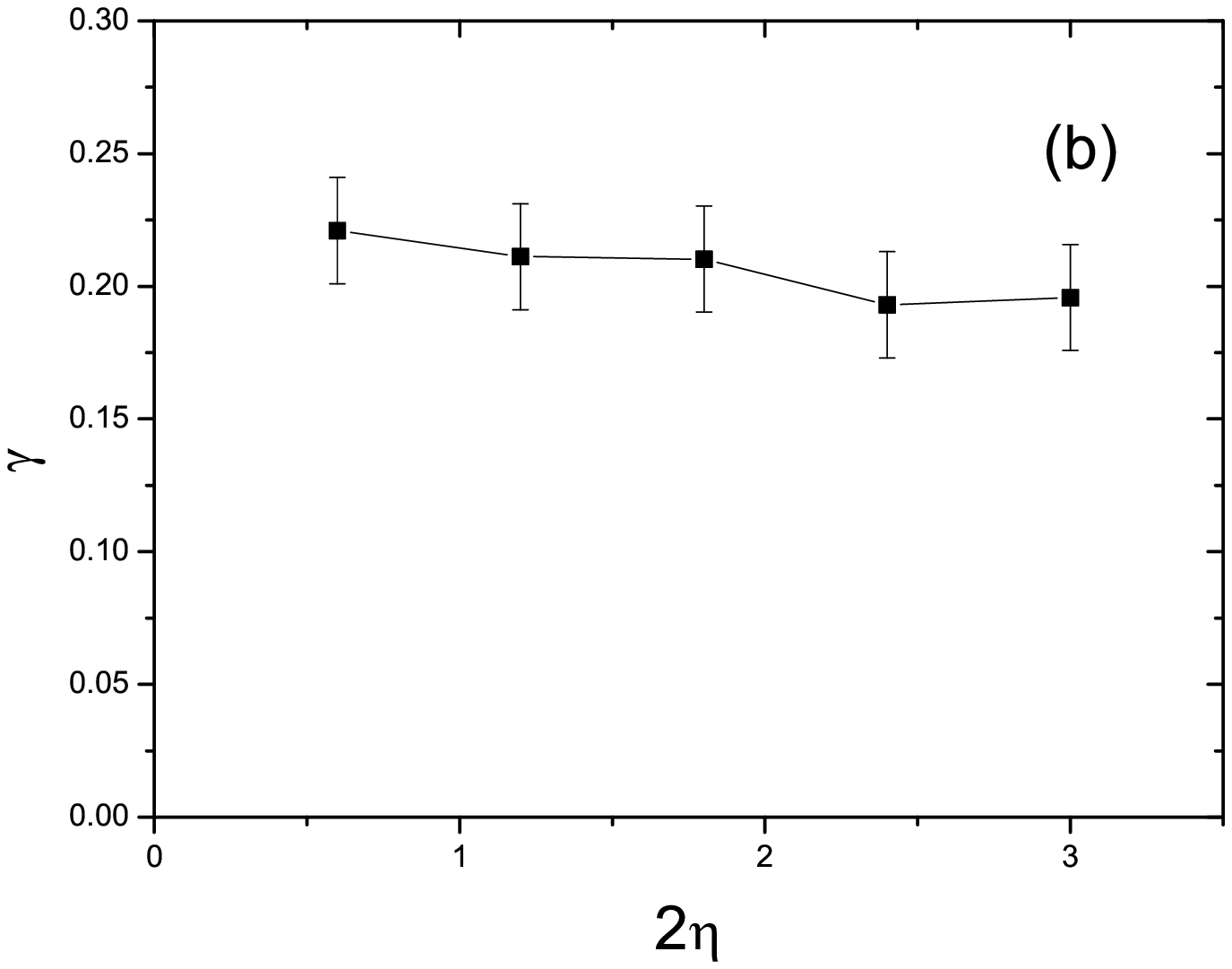}
\caption{(a)The TEE of the state RVB-I as a function of the
subsystem size. Here $L_{A}$ is the linear scale of region A in
Fig.1. The total system size is fixed at $16\times16$. (b)The TEE as
a function of the gap parameter $\eta$. The total system size is
fixed at $12\times12$ and the subsystem size is fixed at $L_{A}=3$.
} \label{fig4}
\end{figure}

\begin{figure}[h!]
\includegraphics[width=8cm,angle=0]{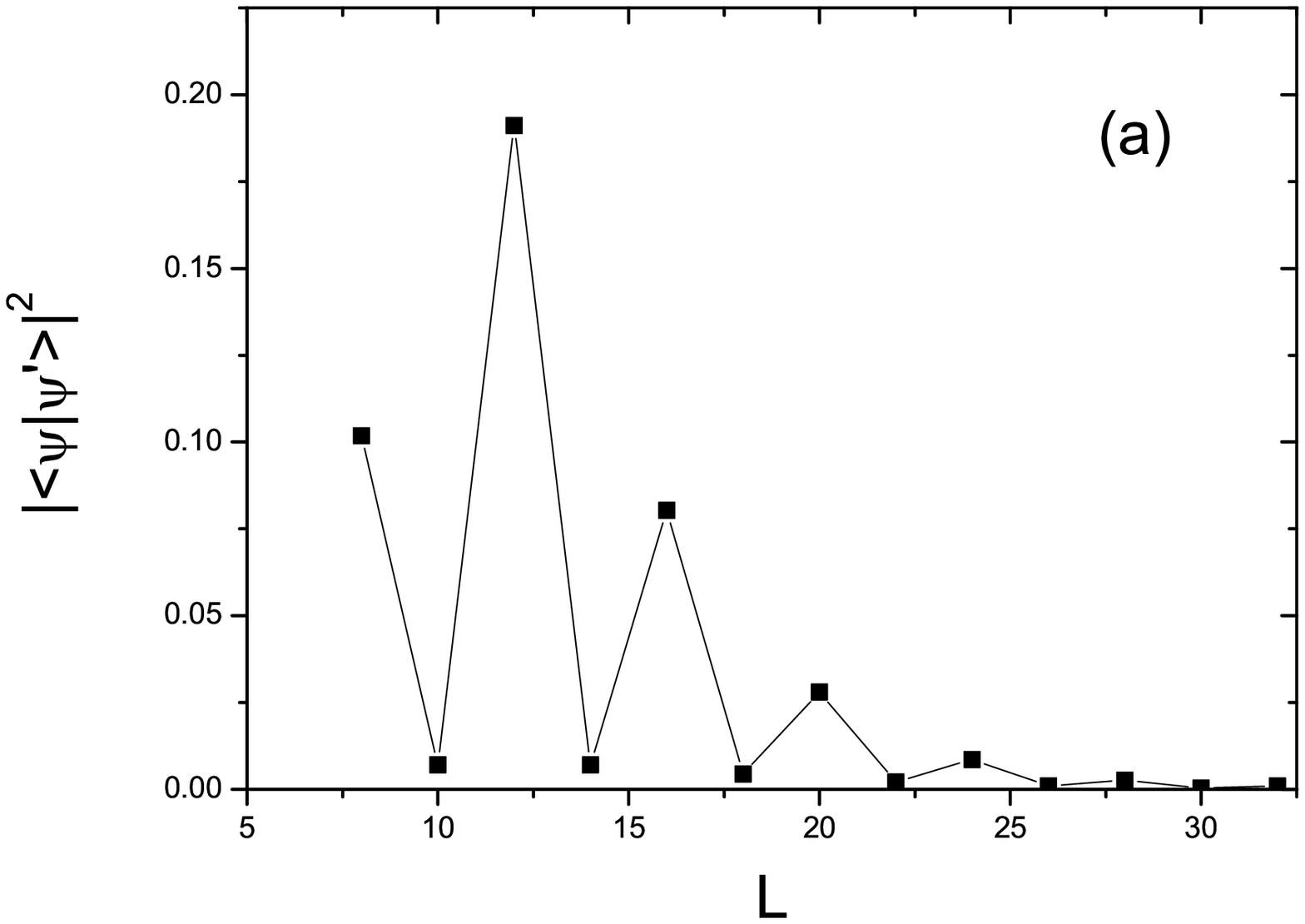}
\includegraphics[width=8cm,angle=0]{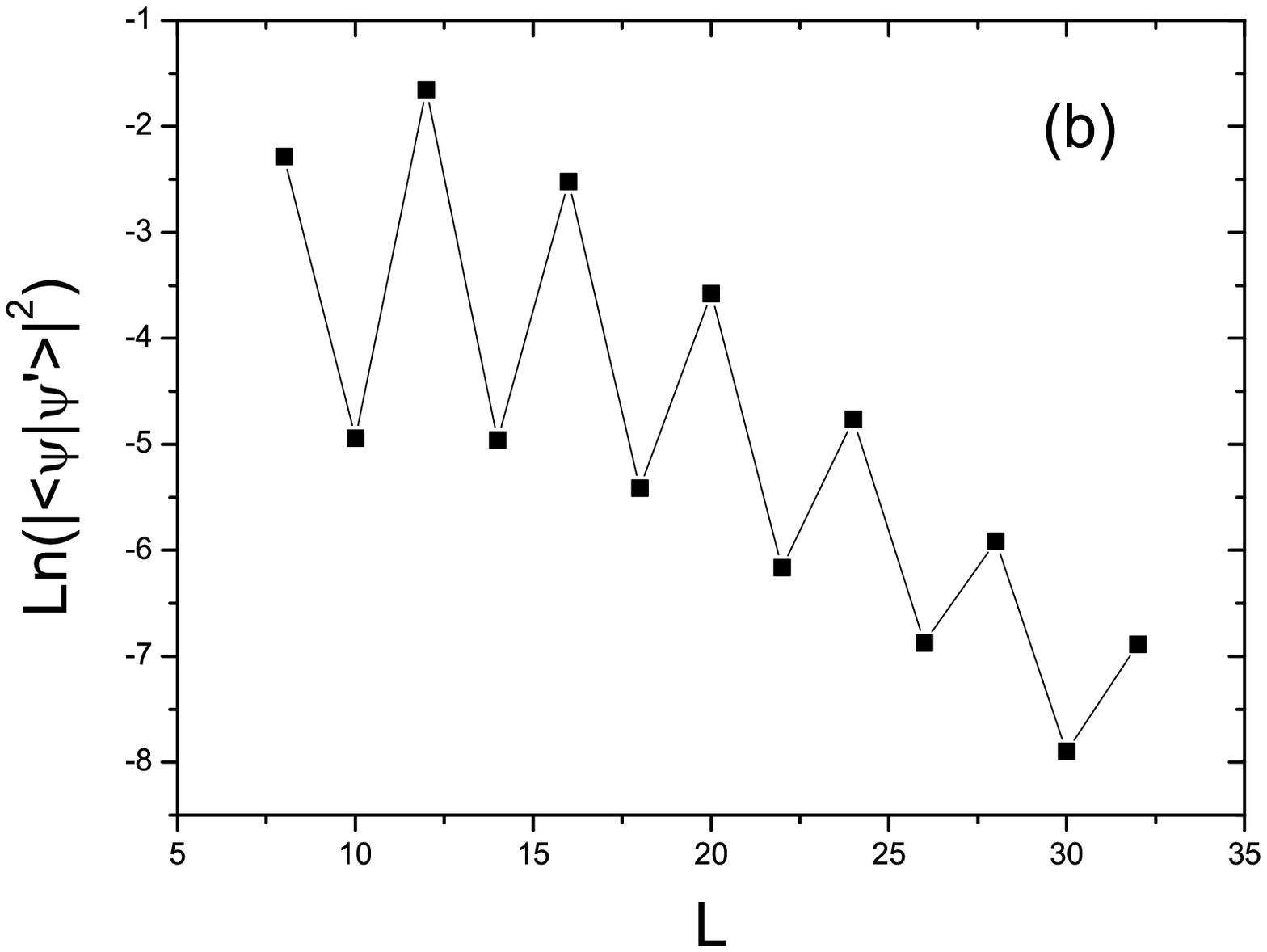}
\caption{The overlap of the wave functions with different number of
inserted visons in the holes of a torus. (a) in linear scale (b) in
logarithmic scale. The decay rate of the overlap is proportional to
the inverse of the vison gap\cite{Arun}. } \label{fig5}
\end{figure}

As a further check of the above result, we perform a test of
topological degeneracy on the state RVB-I. A spin liquid state with
$Z_{2}$ topological order should exhibit four-fold topological
degeneracy on a torus. The four degenerate states differ from each
other in the number of trapped visons in the two holes of the torus.
To detect such degeneracy, we evaluate the overlap of the wave
functions with different number of inserted visons in the holes of a
finite torus. In a topological ordered state, such overlaps should
vanish in the thermodynamic limit. The decay rate of the overlap
provides us an estimate of the size of the vison gap\cite{Arun}. A
nonzero vison gap is the defining property of a spin liquid state
with $Z_{2}$ topological order. It should be noted that the vison
gap has nothing to do with the spin gap. In fact, a spin liquid
state can have a zero vison gap even when its spin correlation is
extremely short ranged. A well known example of this kind is the
short range RVB state defined on the square lattice, for which the
vison gap vanishes as a result of the bipartite nature of the
lattice\cite{RK}.

For the Gutzwiller projected RVB state, inserting a $Z_{2}$ vison in
a hole of torus is equivalent to changing the boundary condition of
the RVB amplitude around that hole from periodic to anti-periodic or
vice versa. For convenience's sake, we choose to evaluate the
overlap between the RVB state with periodic-antiperiodic boundary
condition and that with antiperiodic-periodic boundary condition. In
the calculation we have set $\eta=1.5$ as in Ref.\cite{Vishwanath},
which corresponds to a spin correlation length as short as $1.3$
lattice constant. The result of the overlap is shown in Fig.5. It
fluctuates with the system size and decay only very slowly if
compared with the decay rate of the spin correlation function. This
would imply an extremely small vison gap, if it does exist at all.
We thus find the result of the topological degeneracy is consistent
with result of TEE and both indicate that the topological order in
the state RVB-I is very weak, if it exist at all. The result in
Fig.5 also indicates that to get a converged result for TEE, one
should use subsystem with size much larger than $20\times20$, which
is of course unrealistic.

Thus for the spin liquid state RVB-I, the effective field theory and
the Gutzwiller projected wave function predict totally different
topological properties. To understand the origin of such a
discrepancy, we note that it is previously found that the Marshall
sign rule in the RVB wave function can result in the suppression of
the vison gap and can lead to the absence of topological
degeneracy\cite{Li1,Li2,Li3}. For similar reasons, the approximate
Marshall sign structure in the state RVB-I will definitely enhance
the overlap integral and thus reduce the vison gap.

There is another reason for the suppression of vison gap in the
state RVB-I. As is well known, the $Z_{2}$ topological
classification of a RVB state is only meaningful when the RVB
amplitude is short ranged. However, the RVB amplitude for the state
RVB-I is long ranged, even though its spin correlation is extremely
short ranged. This is because the integrand
$\frac{\Delta_{\mathrm{k}}}{\xi_{\mathrm{k}}+\sqrt{\xi_{\mathrm{k}}^{2}+|\Delta_{\mathrm{k}}|^{2}}}$
in Eq.(16) is singular when $|\Delta_{\mathrm{k}}|=0$ and
$\xi_{\mathrm{k}}<0$. From the self-consistent equation $\langle
c_{i\downarrow}c_{i\uparrow}\rangle=0$, we have
\begin{equation}
\sum_{\mathrm{k}}\frac{\Delta_{\mathrm{k}}}{\sqrt{\xi_{\mathrm{k}}^{2}+|\Delta_{\mathrm{k}}|^{2}}}=0.
\end{equation}
It can be easily shown that to satisfy this equation
$\Delta_{\mathrm{k}}$ must cross zero between the $\Gamma=(0,0)$ and
the $\mathrm{M}=(0,\pi)$ point in the Brillouin zone. Thus the gap
node must occur inside the Fermi surface and RVB amplitude of the
state RVB-I can not be short ranged, although the spin excitation
spectrum is fully gapped. Similar situation also occurs in a
$p+ip'$-wave BCS superconductor, in which the gap node at the
$\Gamma$ point forces the wave function of the Copper pair to have a
long range tail, although the Fermi surface is fully gapped by the
complex pairing\cite{Green}.

\section{V. The topological properties of the state RVB-II}
From the results of the last section, we see the Gutzwiller
projected RVB state may have different topological property from
that predicted by the corresponding effective field theory, even if
the gauge symmetry is broken to $Z_{2}$ and the spin excitation is
fully gapped. More specifically, we find both the Marshall sign rule
and a long ranged RVB amplitude can act to suppress the vison gap.
Following this line of reasoning, we expect that the $Z_{2}$
topological order predicted by the effective field theory should be
robust in a Gutzwiller projected RVB state when it has a short
ranged RVB amplitude and a frustrated sign.

To illustrate this point, we consider a spin liquid state defined on
the triangular lattice. It has the following mean field ansatz
\begin{eqnarray}
U_{i,i+a_{1}}&=&-\tau^{1}\nonumber\\
U_{i,i+a_{2}}&=&-(-1)^{i_{1}}\tau^{1}\nonumber\\
U_{i,i+a_{3}}&=&-(-1)^{i_{1}}\tau^{1}\nonumber\\
U_{i,i}&=&-\mu\tau^{3}.
\end{eqnarray}
Here $a_{1,2,3}$ are the three elementary translation vectors on a
triangular lattice, $i_{1}$ is the coordinate of a lattice site in
the $a_{1}$ direction. The sign of the pairing terms are such that
around each elementary rhombic plaquette of the triangular lattice
the product of the pairing term is negative. This ansatz is
illustrated in Fig.6.
\begin{figure}[h!]
\includegraphics[width=9cm,angle=0]{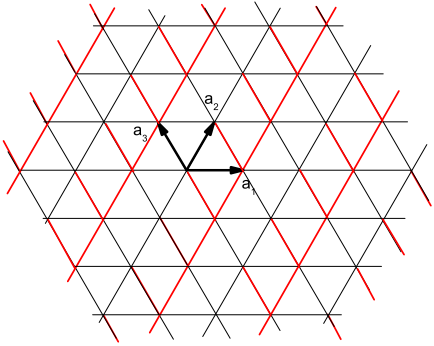}
\caption{An illustration of the triangular lattice and the mean
field ansatz Eq.(18). The pairing term is positive on the red bonds
and negative on the black bonds.\cite{Sorella}} \label{fig6}
\end{figure}

This state is first studied in Ref.\cite{Sorella}. When $\mu
\rightarrow -\infty$, the Gutzwiller projected RVB state derived
from the above ansatz evolves into the short range RVB state
proposed by Anderson on triangular lattice\cite{Anderson}. On the
other hand, when $\mu=0$, it becomes a rather good variational
ground state of the Heisenberg model on triangular lattice and
possesses Dirac type spinon excitation at low energy. The RVB state
for general value of $\mu$ can be viewed as an interpolation between
these two limits.

For nonzero value of $\mu$, the gauge symmetry of the ansatz is
broken to $Z_{2}$ and the spin excitation is fully gapped. At the
same time, the RVB amplitude is short ranged and the sign of the
wave function is frustrated. So we have every reason to expect a
robust $Z_{2}$ topological order in this state.

\begin{figure}[h!]
\includegraphics[width=8cm,angle=0]{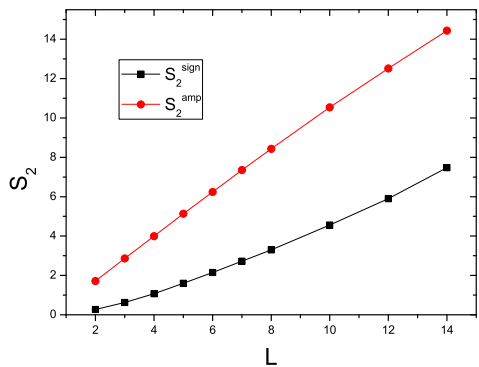}
\caption{The contribution from the sign and amplitude of the wave
function to the Renyi entanglement entropy in the state RVB-II. The
subsystem is now a $L\times L$ rhombic region on the triangular
lattice and the total system size is fixed at $20\times20$. }
\label{fig7}
\end{figure}

\begin{figure}[h!]
\includegraphics[width=8cm,angle=0]{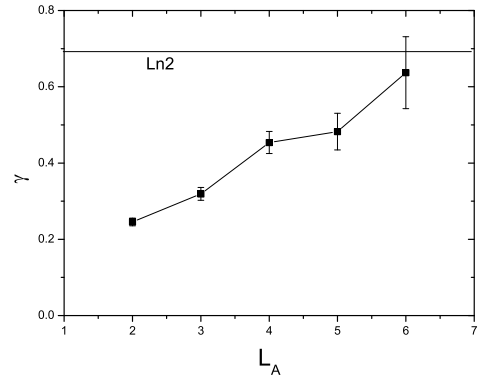}
\caption{The TEE of the state RVB-II as a function of the subsystem
size. Here $L_{A}$ is the linear scale of a rhombic region on
triangular lattice. The total system size is fixed at $20\times20$.
} \label{fig8}
\end{figure}

In the calculation we set $\mu=-30$ to produce a rather short ranged
RVB amplitude. Unlike the state RVB-I, the state RVB-II has a fully
frustrated sign. We find the contribution from the sign now makes up
a much larger portion in the total entanglement entropy. This is
illustrated in Fig.7. The result for TEE is shown in Fig.8. The TEE
is seen to converge steadily to the expected value of $\ln2$ for a
spin liquid with $Z_{2}$ topological order.

In a previous study\cite{Vishwanath1}, it is found that in a
critical RVB state the logarithmic correction to the area law is
mainly contributed by the sign of the wave function. To see how the
sign of the wave function would contribute to TEE, we have plotted
in Fig.9 the contribution to TEE from the sign and the amplitude of
the wave function separately. A key observation is that for both the
state RVB-I and RVB-II, the contribution from the sign and that from
the amplitude always have opposite signs. It is always the sign,
rather than the amplitude of wave function that generate a positive
contribute to TEE. It is also interesting to note that although
$S_{2}^{\mathrm{amp}}>S_{2}^{\mathrm{sign}}$ in both RVB-I and
RVB-II, it is the contribution from the sign that dominate the TEE.
This indicate that the topological property of the RVB state is
mainly determined by its sign structure.

\begin{figure}[h!]
\includegraphics[width=8cm,angle=0]{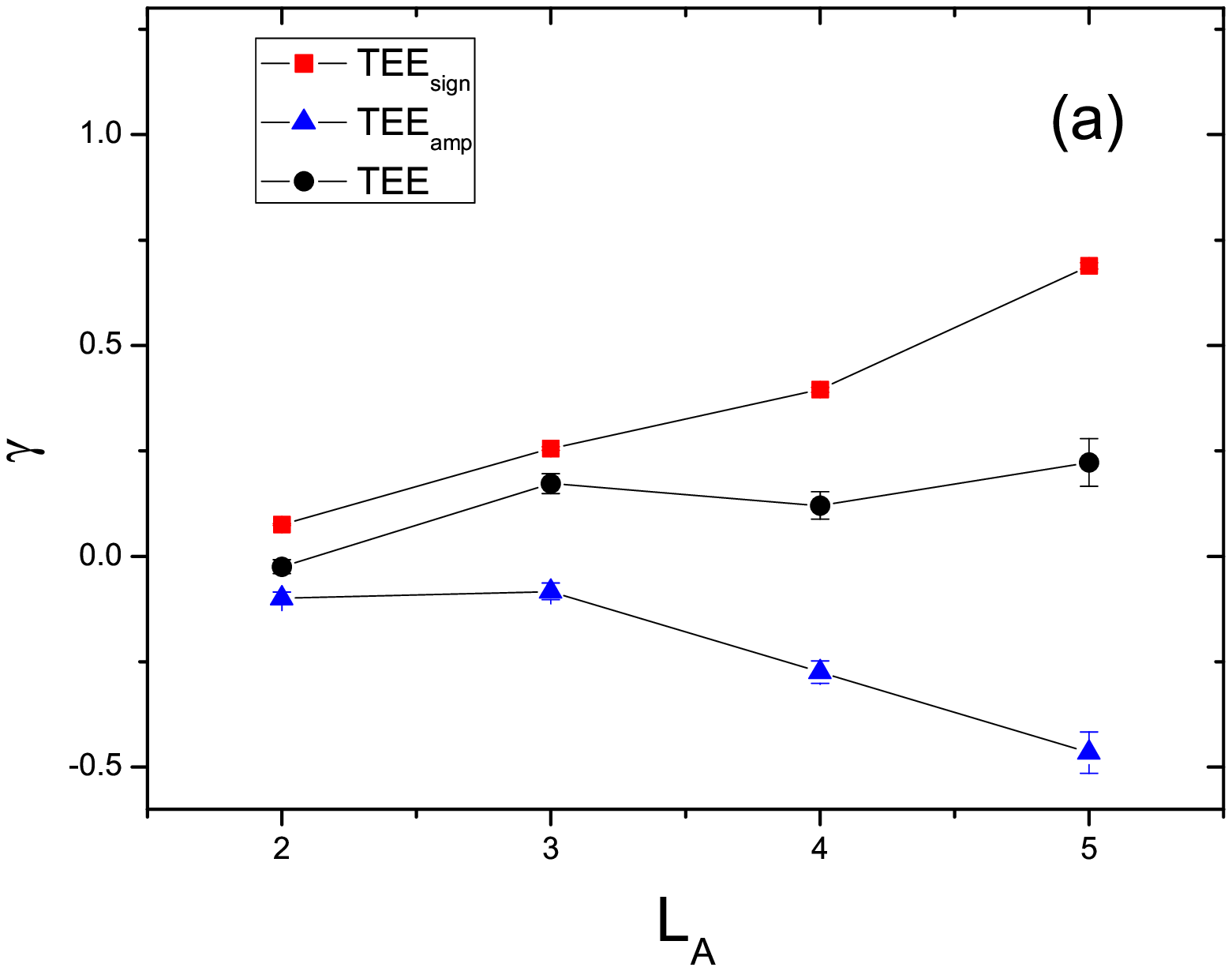}
\includegraphics[width=8cm,angle=0]{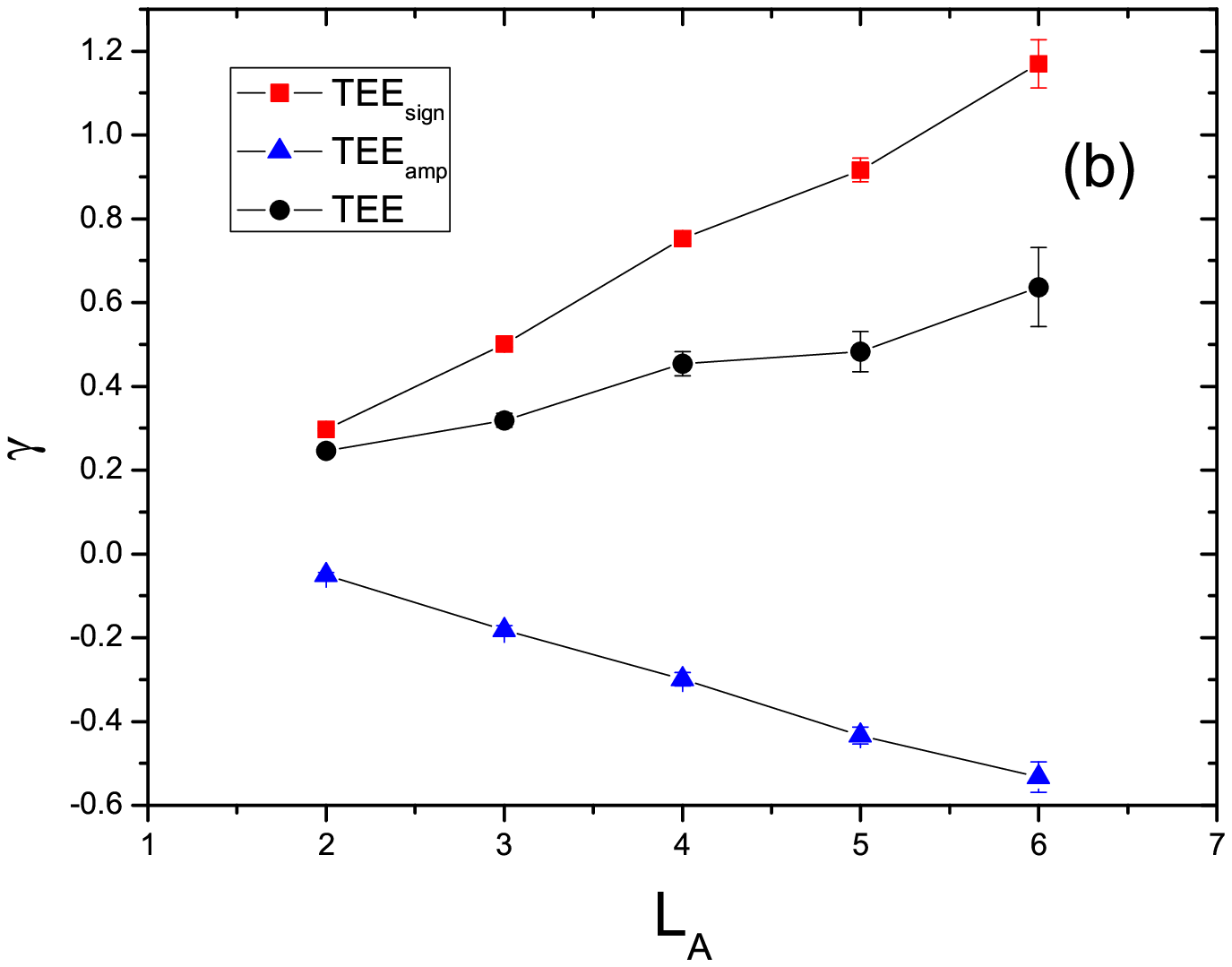}
\caption{The contribution of the sign and the amplitude of the wave
function to the TEE in the state RVB-I(a) and RVB-II(b). The total
system size is fixed at $16\times16$ and in (a) and is fixed at
$20\times20$ in (b). } \label{fig9}
\end{figure}

\section{VI. Discussions}

In this paper, we have evaluated the TEE for two Gutzwiller
projected Fermionic RVB states. Both states are derived from mean
field ansatz with a $Z_{2}$ gauge structure and a full gap to spin
excitations. Following effective field theory arguments, both states
should exhibit $Z_{2}$ topological order. However, from the result
of TEE we find the two states have different topological properties.
With new tricks to improve the convergence of TEE, we are able to
show that the spin liquid state RVB-I studied in
Ref.\cite{Vishwanath} does not support $Z_{2}$ topological order,
while the spin liquid state RVB-II proposed in Ref.\cite{Sorella}
exhibits robust  $Z_{2}$ topological order.

These results indicate that something beyond the effective field
theory is important to the topological property of a RVB state. We
find both a long ranged RVB amplitude and an unfrustrated sign in
the RVB wave function can act to suppress the vison gap and thus the
topological order in a RVB state. It is important to note that for
the Fermionic RVB state, the RVB amplitude need not be short ranged
even if the system has a large spin gap and that its spin
correlation is extremely short ranged. The gap node deep inside the
Fermi surface can have profound effect on the long wavelength
behavior of the system. This is markedly different from a Bosonic
RVB state, for which a full spin gap always corresponds to a short
ranged RVB amplitude.

We have also tested topological degeneracy on the state RVB-I. We
find the result implies a very small or even zero vison gap for the
state RVB-I, which is consistent with the conclusion we get from the
result of TEE. To our knowledge, this is for the first time that the
topological property of a RVB state is studied from both the TEE and
the topological degeneracy perspective with consistent conclusions
reached. We also find that it is the sign, rather than the amplitude
of the wave function that dominates the TEE and is responsible for a
positive value of TEE in the Gutzwiller projected RVB state. Since a
positive TEE is the signature of nonlocal entanglement in the
system, we find the nonlocal entanglement in a RVB state is mainly
encoded in its sign structure. This is consistent with previous
studies in which it is found that a Marshall sign rule is at the
origin for the absence of topological order in certain spin
liquids\cite{Li1,Li2,Sorella1}.

From the effective field theory point of view, a $Z_{2}$ gauge
structure and a fully gapped spin excitation spectrum is enough to
guarantee the stability of the saddle point against gauge
fluctuation up to Gaussian level. Since Gutzwiller projection just
amounts to gauge averaging on the saddle point, we expect the
Gutzwiller projected RVB state to have the same topological property
as the saddle point action of effective field theory, if the
fluctuation effect beyond the Gaussian level can be neglected in the
long wavelength limit. Our results show that this is not always
true. Thus, certain singular fluctuation mode beyond the Gaussian
level is important for the topological property of RVB state. It is
an interesting problem to study how such singular gauge mode are
related to the emergence of the Marshall sign rule structure of the
Gutzwiller projected wave function and to the long ranged nature of
the RVB amplitude.

In conclusion, from our results it is clear that the effective field
theory at the Gaussian level misses two pieces of information that
is important for the topological property of a RVB state. The first
one is the sign structure of the Gutzwiller projected wave function.
The second one is the range of the RVB amplitude of the RVB state.
The effective field theory can predict the topological property of a
RVB state reliably only when both of these information are accounted
for.

This work is supported by NSFC Grant No. 10774187, No. 11034012 and
National Basic Research Program of China No. 2010CB923004.

\end{document}